\documentclass[twocolumn,showpacs,preprintnumbers]{revtex4}
\def\pref#1{(\ref{#1})}
\usepackage{graphicx} 
\usepackage{amssymb}
\usepackage{amsmath}

\begin{document}


\title{Polaritonic spectroscopy of intersubband transitions}

\date{\today}

\author{Y. Todorov}\email{yanko.todorov@univ-paris-diderot.fr}
\affiliation{Univ. Paris Diderot, Sorbonne Paris Cit{\'e}, Laboratoire Mat{\'e}riaux et Ph{\'e}nom{\`e}nes Quantiques, UMR7162, 75013 Paris, France}
\author{L. Tosetto}
\affiliation{Univ. Paris Diderot, Sorbonne Paris Cit{\'e}, Laboratoire Mat{\'e}riaux et Ph{\'e}nom{\`e}nes Quantiques, UMR7162, 75013 Paris, France}
\author{A. M. Andrews}
\affiliation{Solid State Electronics Institute TU Wien, Floragasse
7, A-1040 Vienna, Austria}
\author{A. Delteil}
\affiliation{Univ. Paris Diderot, Sorbonne Paris Cit{\'e}, Laboratoire Mat{\'e}riaux et Ph{\'e}nom{\`e}nes Quantiques, UMR7162, 75013 Paris, France}
\author{A. Vasanelli}
\affiliation{Univ. Paris Diderot, Sorbonne Paris Cit{\'e}, Laboratoire Mat{\'e}riaux et Ph{\'e}nom{\`e}nes Quantiques, UMR7162, 75013 Paris, France}
\author{G. Strasser}
\affiliation{Solid State Electronics Institute TU Wien, Floragasse
7, A-1040 Vienna, Austria}
\author{C. Sirtori}
\affiliation{Univ. Paris Diderot, Sorbonne Paris Cit{\'e}, Laboratoire Mat{\'e}riaux et Ph{\'e}nom{\`e}nes Quantiques, UMR7162, 75013 Paris, France}

\begin{abstract}
We report on an extensive experimental study of intersubband excitations in the THz range arising from the coupling between a quantum well and a zero-dimensional metal-metal microcavities. Because of the conceptual simplicity of the resonators we obtain an extremely predictable and controllable system to investigate light-matter interaction. The experimental data is modelled by combining a quantum mechanical approach with an effective medium electromagnetic simulation that allows us to take into account the losses of the system. By comparing our modelling with the data we are able to retrieve microscopic information, such as the electronic populations on different subbands as a function of the temperature. Our modelling approach sets the base of a designer tool for intersubband light-matter coupled systems. 
\end{abstract}
\maketitle

\section{Introduction}
The ability of engineering semiconductor devices at the nanometer scale, by modifying their quantum properties, has been an essential ingredient that has allowed the continuous impressive technological development of electronics and optoelectronics. A prominent example is provided by the emergence of new laser sources in the Mid-IR and THz regions, the so called Quantum Cascade Lasers \cite{Faist_QCL_1994}. In these structures the electronic transport, under the application of an electrical bias is carefully designed in order to obtain a redistribution of the carriers on the different energy levels and finally population inversion between subbands of multiple-coupled semiconductor quantum wells. 

The engineering of the light-matter interaction in an intersubband (ISB) system resides somehow on the opposite limit, as it can be enhanced by achieving a large electronic population on the fundamental subband of a quantum well inserted in a microcavity with a very small volume. In this case the electromagnetic response of the system is dominated by collective plasmonic excitations of the 2D electron gaz \citep{Ando_Fowler_Stern_1982}. When combined with an optical microcavity, these systems enter the strong coupling regime, in which the interaction between the electromagnetic and the electronic collective modes yield new mixed intersubband polariton states \citep{Dini_2003}. The splitting between the mixed states, also called Rabi splitting, is a direct measure of the light-matter coupling strength. The latter can be increased through doping by increasing the number of electrons participating to the interaction. This enabled to reach the ultra-strong coupling regime, where the Rabi splitting becomes comparable with the energy of the ISB plasmon mode \cite{Ciuti_PhysRevB_2005}. This regime has been intensively studied and reported by several groups, even at THz frequencies and up to room temperature \cite{Todorov_PRL_2009,Anappara_2009a,PJouy_2011,Geiser_2012}.     

The aim of this work is to interpret the spectroscopic features of the polariton states as a function of the temperature in order to extract microscopic information on intersubband transitions. In this respect we will show that measurements of the Rabi splitting, $2\Omega_R$, permit to follow the electronic distribution on the different subbands. Indeed, $2\Omega_R$ has a straightforward dependence on the subband population and the corresponding transition oscillator strengths. This is clearly illustrated in the formula of the Rabi splitting for a single ISB excitation between the fundamental and second subband, coupled with the lowest order $\mathrm{TM}_0$ mode of a double-metal waveguide \cite{Andreani_2003, Todorov_PRB_2012}:

\begin{equation}\label{1_Splitting}
2\Omega_R = \sqrt{\frac{f_{12}e^2 N_{\mathrm{QW}}(N_1-N_2)}{\varepsilon \varepsilon_0 m^\ast L_\mathrm{cav}}}
\end{equation}

Here $e$ is the electron charge and $\varepsilon_0$ is the vacuum permittivity. The effective electron mass $m^\ast$ and the dielectric constant of the semiconductor core $\varepsilon$ are well known material parameters.  The parameter $L_\mathrm{cav}$ is the cavity thickness, $N_{\mathrm{QW}}$ is the number of quantum wells contained in the cavity, $N_1/N_2$ is the electronic population of the fundamental/second subband, and $f_{12}$ is the transition oscillator strength that depends on the overlap between wave-functions. Eq.\pref{1_Splitting} indicates that the populations, $N_1$ and $N_2$ can be inferred from the study of the splitting $2\Omega_R$ as a function of the temperature, by applying the Fermi-Dirac statistics.

\begin{figure}
\includegraphics[scale=0.42]{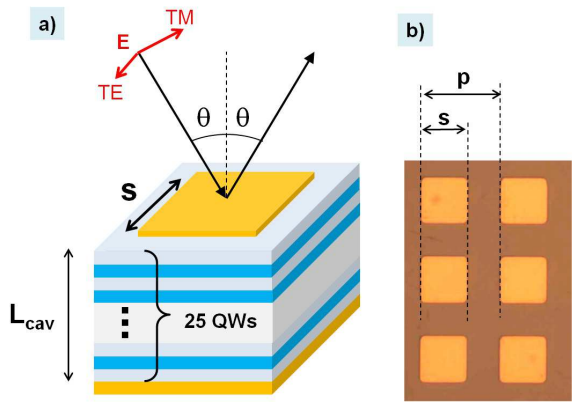}
\caption {(a) Schematic of an individual metal-metal patch cavity with thickness $L_\mathrm{cav}$, containing  25 $\mathrm{GaAs/AlGaAs}$ quantum wells. The structure is probed in reflectivity measurements, with an incident angle $\theta$, in both TM ($p$-) and TE($o$-) polarization. (b) Microscope picture of the top surface of the sample, containing an  array of patch cavities. The size of the patches is $s$, and the period of the array is $p$.} \label{Fig1}
\end{figure}

Our study is conducted with 0D double-metal dispersionless square-patch resonators, as described in Fig.\ref{Fig1}, which operate on standing wave-like $\mathrm{TM}_0$ modes \citep{Todorov_Opex_2010}. The conceptual simplicity of this geometry reduces considerably the photonic degrees of freedom, allowing the direct study of the electronic component of the polariton states. Indeed, these cavities feature a flat photonic dispersion, and the resonant frequency is independent from the incident angle of the probing beam. Moreover, the electric field that couples to the intersubband polarization is homogeneous along the growth axis of the quantum wells. The only parameter that governs the coupling is the cavity mode detuning from the ISB resonance, which in turn is deterministically set by the size $s$ of the patch \citep{Todorov_Opex_2010}. 

Although a more complex fabrication procedure, the metallic patch cavities have several advantages with respect to the multi-pass absorption geometry that is commonly used to study the intersubband  transitions \cite{Book_Helm}.  First of all they allow performing intersubband transition spectroscopy with light propagating normal to the wafer surface and therefore with a much larger surface exposed to the impinging. Moreover, in this geometry the light does not traverse the semiconductor substrate and avoids all the spurious absorptions in the bulk material. We should nevertheless carry in mind that Eq.\pref{1_Splitting} describes an ideal system, which does not take into account all the details of a typical experimental sample.  To correctly deduce the electronic distribution from the experimentally measured Rabi splitting one needs to consider several other complications, which are typically not included in an purely energy preserving Hamiltonian description: 

i) An important aspect is the dissipation in the system, which causes finite linewidths both for the cavity modes and intersubband resonances. Therefore the measured Rabi splitting is actually reduced with respect to values predicted by Eq.\pref{1_Splitting} \citep{Savona1995733}. 

ii) In highly doped samples, several ISB resonances can be active at the same time. Their interaction causes a redistribution of  the oscillator strength \cite{Warburton_1998}. Therefore we need a multi-transition generalization of Eq.\pref{1_Splitting}. 

iii)  For the THz region, where large quantum wells are employed, the Hartree potential induces important correction to the heterostructure potential. Therefore energy levels and wave functions should be obtained by self-consistent Schr\"{o}dinger-Poisson calculation. This should be taken into account for the correct application of the Fermi-Dirac statistics.

iv) The presence of metal-dielectric junctions gives rise to a charge transfer, which changes the carrier concentration of the quantum wells close to the junction \citep{Book_Sze}. This effect is important for the correct estimate of the number of charged quantum wells $N_{\mathrm{QW}}$.

The structures that we  investigate, already reported in Ref. \cite{Todorov_PRL2010}, are detailed in the subsequent sections. Our study is based on experimental results that were not presented in that reference, and on an analysis that takes into account the afore mentioned points i)-iv). 

The paper is organized as follows. In part \ref{PartII} we present the experimental system and we briefly review the photonic confinement in our structures. Part \ref{PartIII} deals with an extensive study of the intersubband polaritons at different temperatures and  for different cavity detuning with respect to the ISB resonances. We show how the information collected in these experiments, in combination with a  self-consistent simulation of  the heterostructure potential allows to infer the electronic population of different subbands as a function of the temperature.  In particular,  we interpret our absorption data at THz frequency as a function of the temperature \cite{Todorov_PRL2010}, and show that they actually display the transfer of oscillator strength \cite{Warburton_1998}.  

\section{Experimental system}\label{PartII}

\subsection{Microcavity}

The photonic structure  employed in our experiments is summarized in Fig.\ref{Fig1} that sketches a single square-patch double-metal microcavity containing 25 $\mathrm{GaAs}/\mathrm{AlGaAs}$ modulation doped quantum wells. The characterization of the quantum wells will be presented in details in the next section. The sample contains a dense array of such micro-cavities, as illustrated in Fig.\ref{Fig1}(b), in which the period of the array $p$ is kept close to the square size $s$, so to optimize the coupling efficiency with the incident radiation \citep{Todorov_Opex_2010}. The overall thickness of the core region between the two metals is $L_\mathrm{cav}=$1.5$\mu$m.

\begin{figure}
\includegraphics[scale=0.45]{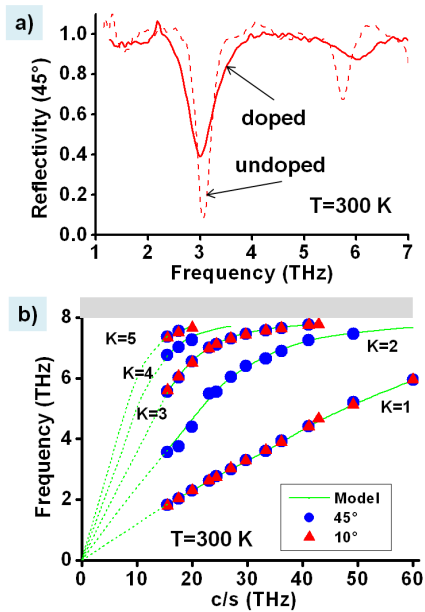}
\caption {(a) Typical room temperature reflectivity spectra for two samples with an array of patches with $s=$12$\mu$m. Dashed line: undoped reference sample. Solid line: sample with doped quantum wells, but with the same overall thickness ($L_\mathrm{cav}$=1.5$\mu$m) as the reference. (b) Resonance frequency of the doped sample, measured at 300 K, as a function of the inverse square size multiplied by the speed of light $(c/s)$. $K$ denotes the order of the resonant mode, and the grey area corresponds to the optical phonon band.} \label{Fig2}
\end{figure}

The structure is tested by varying the incident angles $\theta$ (as defined in Fig.\ref{Fig1}(a)) and the electric field polarization of  the incident beam. Characteristic reflectivity spectra for a cavity made of an undoped GaAs thick layer (dotted line) and for a cavity with quantum wells (solid line) are shown in Fig.\ref{Fig2}(a). Data are taken at room temperature (300 K) and for an angle of incidence of $\theta = 45^\circ$. For both samples the patch sizes are equal to $s$ = 12$\mu$m. These spectra display dips at the energies of the resonant cavity modes of the structure. For both structures we observe two resonances which correspond to the excitation of the first two $\mathrm{TM}_0$-like standing wave modes \citep{Todorov_Opex_2010}, around 3 THz and 6 THz. In the case of the cavity with quantum wells, at room temperature, there is no clear signature of the ISB resonances.  However the effect the electron gas is still visible in the data of  Fig.\ref{Fig2}(a) through the considerable enlargement of the cavity peaks for the doped sample. Indeed, at high temperature, the electrons provided by the Si-donors can scatter in the confined and continuum states of the heterostructure by giving rise to sizeable free carrier losses.  

In Fig.\ref{Fig2}(b) we summarize the resonant frequencies of measured doped structure at $T=$ 300 K as a function of the inverse square size multiplied by the speed of light: $c/s$. Measurements for incident angles $\theta = 45^\circ$ (circles) and $\theta = 10^\circ$ (triangles) are both reported. At low frequencies we recover the typical linear dispersion of the $\mathrm{TM}_0$ guided mode, $\nu_K = Kc/n_{eff}s$, with $K$ the order of the lateral standing wave and $n_{eff}$ the effective index. At higher frequencies, the dispersion becomes non-linear due to coupling with the optical phonon of the semiconductor, which results in a strong increase of the effective index. The dispersion is independent from the incident angle, however close to normal incidence ($\theta = 10^\circ$) only the resonances with odd $K$ are excited due to the selection rule of the grating \citep{Todorov_Opex_2010}. The position of the resonances is well described by a modal method model for a linear grating \citep{Todorov_JOSAA_2007}, based on the scattering matrix formalism \cite{Collin_S_2001}. Indeed, the modes of the square patches of the structure, that couple better to the incident beam, have a dipolar charge distribution \citep{Todorov_Opex_2010}, and these modes are identical to the resonances of the double-metal lamellar grating structure. We will therefore use this formalism for the data modelling that is described in part \ref{PartIII}.

\subsection{Quantum well slab}

The semiconductor region is a $\mathrm{GaAs}/\mathrm{Al_{0.15}Ga_{0.75}As}$ heterostructure made of 32nm thick quantum wells (QWs) and 20nm thick barriers. $X$-ray analyses indicate a slightly lower average Al content in the barriers 12$\%$ than the nominal 15$\%$. The structure is $\delta$-doped in the each barrier, 5nm away from the adjacent quantum well, with a nominal value of $2 \times 10^{11} \mathrm{cm}^{-2}$ of the sheet carrier concentration. 

\begin{figure}
\includegraphics[scale=0.33]{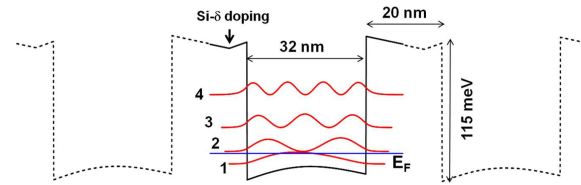}
\caption {Typical conduction band profile of the periodic quantum well stack, including self-consistent Hartree corrections. We have highlighted a single quantum well together with the moduli squared of the first 4 wavefunctions, as well as the Fermi-level energy, $E_F$.} \label{Fig3}
\end{figure}

\begin{figure}
\includegraphics[scale=0.42]{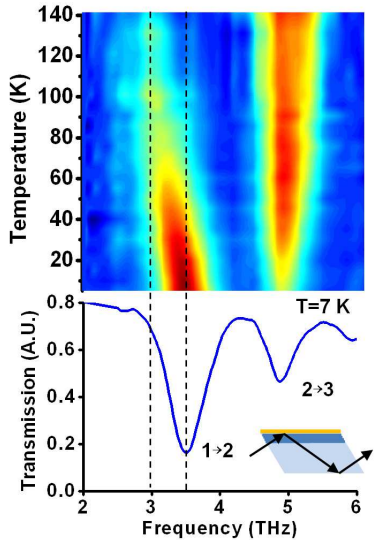}
\caption {(a) Multipass absorption of the quantum well medium, measured as a function of the temperature and frequency. The two features around 3 THz and 5 THz correspond respectively to the 1$\rightarrow$2 and 2$\rightarrow$3 transitions. (b) The absorption spectrum at $T=$7 K. The inset illustrates the experimental configuration.} \label{Fig4}
\end{figure}

In Fig.\ref{Fig3} we show a typical conduction band profile obtained with Poisson-Schr\"{o}dinger solver with periodic boundary conditions, at low temperature ($T=$7 K), including the Hartree corrections to the quantum well potential. The envelope wave-functions of the confined subbands of the quantum well are also indicated. Details on these simulations are given in Appendix \ref{App_PS_solver} and in part \ref{PartIII}.  The absorption of the quantum well is measured in a multi-pass configuration with $45^\circ$ polished facet and a gold layer deposited on top, as indicated in Fig.\ref{Fig4} (inset). The results of the experiment are summarized in Fig.\ref{Fig4}, where two absorption peaks are clearly visible from 7 K to 150 K. These peaks correspond to the 1$\rightarrow$2 and 2$\rightarrow$3 transitions of the quantum well of Fig.\ref{Fig3}.  The contour plot of Fig.\ref{Fig4}(a) is obtained by merging multi-pass transmission spectra performed at different temperatures. In the contour plot the red color scale corresponds to a low transmission signal and therefore strong ISB absorption. In the lower part of the figure we report the spectrum at $T=$7 K, where the 3.5 THz peak corresponds to the 1$\rightarrow$2 transition, and the 4.8 THz peak is the 2$\rightarrow$3 transition.

As expected from Fermi-Dirac statistics, the intensity of the two peaks have an opposite dependence on temperature: the first 1$\rightarrow$2  transition  is activated at low temperature as electrons accumulate on the ground state of the quantum well and the population difference $N_1-N_2$ is increased. Moreover, a strong blue shift is visible for the 1$\rightarrow$2  transition, due to the depolarization effect \citep{Ando_Fowler_Stern_1982}. The transition 2$\rightarrow$3 follows an opposite behaviour, since the subband 2 is progressively populated as the temperature is risen. The fact that the 2$\rightarrow$3 transition is visible even at low temperature indicates that the Fermi level lies above the second subband edge, meaning that the doping is higher than the nominal value. Further, we shall see  that the ISB absorption of the 2$\rightarrow$3 transition is increased due to the oscillator strength transfer between the first and the second ISB excitation of the quantum well \cite{Warburton_1998}. Namely, this phenomenon will be elucidated and quantified from the study of the splitting induced by the light-matter interaction on each individual transition.

\subsection{Polariton dot measurements}

When the QW medium is inserted in a 0D microcavity the strong interaction between the ISB excitation and the cavity mode gives rise to two polariton states. In our system the spectroscopic features are a function of the size of the resonator $s$, which sets the detuning between the cavity mode and the ISB excitations. 

\begin{figure}
\includegraphics[scale=0.45]{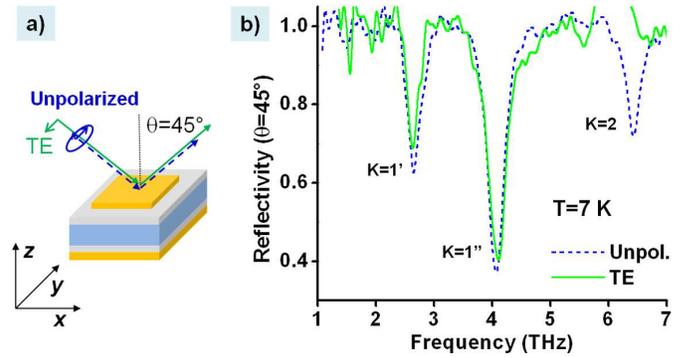}
\caption {(a) Experimental configuration for reflectivity measurements with variable polarization of the incident beam. (b) Low temperature ($T=$7 K) reflectivity spectra performed at an incident angle $\theta=45^\circ$. Solid curve: TE linear polarization. Dashed curve: unpolarized light. The size of the patches is $s$=11.1$\mu$m.} \label{Fig5}
\end{figure}

\begin{figure}
\includegraphics[scale=0.45]{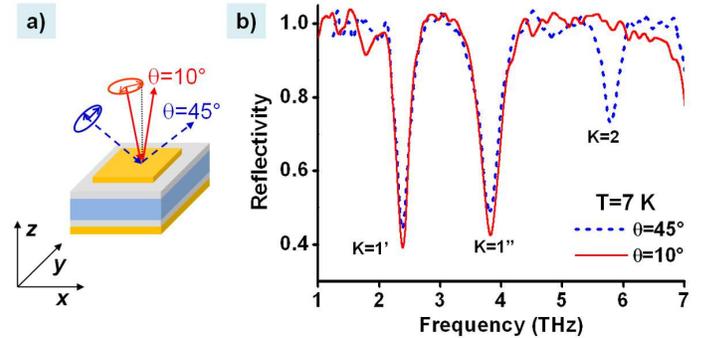}
\caption {(a) Experimental configuration for reflectivity measurements with different incident angles. (b) Low temperature ($T=$7 K) reflectivity spectra performed with un-polarized light, for an angle  $\theta=10^\circ$ (solid line) and $\theta=45^\circ$ (dashed line). The size of the patches is $s$=12.5$\mu$m.} \label{Fig6}
\end{figure}

\begin{figure*}
\includegraphics[scale=0.58]{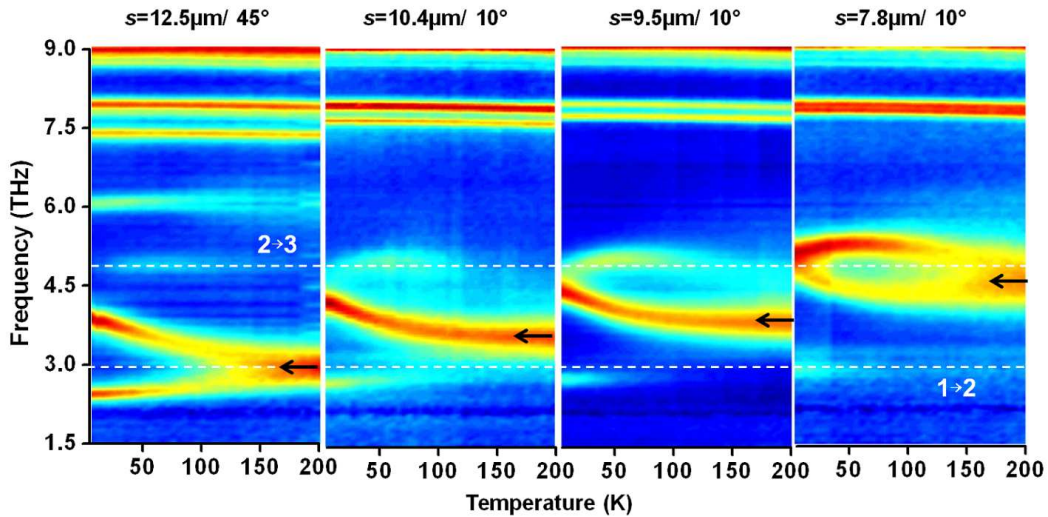}
\caption {Frequency - temperature contour plots of the reflectivity for cavities with different sizes $s$. The resonant frequencies of the cavities (indicated by arrows) have been chosen so that the microcavity modes are tuned between the  1$\rightarrow$2 and 2$\rightarrow$3 ISB transitions. The dashed lines indicate the values of the ISB resonances measured at high temperature, $T>140$ K, according to the data in  Fig.\ref{Fig4}.} \label{Fig7}
\end{figure*}

The polariton splitting due to the coupling between matter excitations and the first order mode of our 0-dimensional cavities is very insensitive to the experimental conditions. This can be seen in the series of different experimental conditions reported in the data of Fig.\ref{Fig5} and Fig.\ref{Fig6}. These experiments are performed at low temperature, $T=$7 K, where the 1$\rightarrow$2 ISB excitation is coupled with the fundamental $K=1$ cavity mode. This mode is twofold degenerate into $\mathrm{TM}_{100}$ and $\mathrm{TM}_{010}$ modes, each oscillating along the two sides of the square stripe (along the $x$- and $y$- direction respectively). Since the two modes have identical coupling with the ISB transition, the polariton splitting is independent from the particular polarization state of the incident beam, as described in Fig.\ref{Fig5}, where we compare two spectra obtained with unpolarised (dotted line) and polarized (solid line) light. If the incident beam is $y$-polarized, then only the $\mathrm{TM}_{010}$ mode is excited. The resulting polariton reflectance spectrum, obtained for an angle of incidence $\theta=45^\circ$ is shown in solid line in Fig.\ref{Fig5}. The size of the patch in this experiment is $s$=11.1$\mu$m. The two polariton peaks are indicated as $K=1^\prime$ and $K=1^{\prime \prime}$. The same experiment repeated with unpolarized light (dashed line) gives rise to practically an identical spectrum (dashed curve in Fig.\ref{Fig5}). As a matter of fact for $\theta=45^\circ$ both $\mathrm{TM}_{100}$ and $\mathrm{TM}_{010}$ are excited and therefore the polariton states $K=1^\prime$ and $K=1^{\prime \prime}$ can be similarly excited for the polarization in the $x$- and $y$- directions. Contrarily, one can see that the second order $\mathrm{TM}_{200}$ mode (indicated as $K=2$) has a different dependence and appears only if a component of the incident electric field  in the $x$- direction is present. This is due to a symmetry selection rule of the even resonances \citep{Todorov_Opex_2010}. 

In Fig.\ref{Fig6} we report the measurements performed with unpolarized light, but at different incident angles:  $\theta=10^\circ$ (solid curve) and $\theta=45^\circ$ (dashed curve). The patch size is $s$=12.5$\mu$m in this case. Once more we can see that the polaritons ($K=1^\prime$ and $K=1^{\prime \prime}$) are unaffected by the angle of incidence, since they belong to the coupling with the first order mode. The $K=2$ resonance disappears at normal incidence due to its symmetry selection rule \citep{Todorov_Opex_2010}.

The above properties of the polariton states, that attest their 0D nature, stem directly from the properties of the $K=1$ resonance, which is dispersionless (independent from the angle of the incidence) and insensitive to the polarization because of the twofold degeneracy arising from the symmetry of the structure. These properties are imprinted on the ISB polarization through the strong interaction with light. We therefore refer to these structures as "polariton dots" \citep{Todorov_PRL2010}.      

Having reduced the light degrees of freedom as much as possible, we can now concentrate on the study of the polariton characteristics as a function of their matter part, i.e. of the ISB excitation. To this end we have measured the temperature evolution of the polaritons for different cavity resonances, as illustrated in Fig.\ref{Fig7}. The four panels in Fig.\ref{Fig7} represent the reflectivity contour plots as a function of the temperature for cavities, in which the first order ($K=1$) resonance is varied between the 1$\rightarrow$2  and 2$\rightarrow$3 ISB transitions. As an indication, the dashed lines in Fig.\ref{Fig7} report the values of the ISB resonances measured at high temperature, $T>140$ K (for graphical simplicity we do not consider here the temperature dependence of the ISB resonances). Note that these values are closest to the ones that one would expect for the same QW without charge effects. In the color code, red corresponds to low reflectance, and blue corresponds to close to unity reflection. The cavity resonance is indicated with an arrow. The first panel (cavity with $s$=12.5$\mu$m) is resonant with the 1$\rightarrow$2 transition, while the fourth ($s$=7.8$\mu$m) is resonant with the 2$\rightarrow$3 transition. The first panel is performed at $45^\circ$ angle of incidence, while the others are performed at $10^\circ$. This is why in the first panel it is visible also the second order resonance ($K=2$) at 6 THz. One can see that as the temperature is lowered below 150 K electrons accumulate preferentially on the first subband, making the ratio $N_1/N_2 >> 1$. The 1$\rightarrow$2 transition is thus activated resulting in a progressively increasing of the polariton splitting for the $s$=12.5$\mu$m cavity or, a growing absorption dip for the off-resonant cases. The 2$\rightarrow$3 transition is most active in the temperature range (50 K $< T <$ 150 K), as qualitatively expected from the Fermi-Dirac statistics. 

The main advantage of the microcavity measurements with respect to multi-pass measurements is the type of spectroscopic data that is related to the population differences between the QW subbands. In a multi-pass experiment, such as the one depicted in Fig.\ref{Fig4}, if a linear absorption regime is assumed, the population difference $N_1-N_2$ is related to the integral of the absorption peak in the spectra \citep{Book_Helm}. The peak area determination requires a careful baseline calibration, which can become difficult in the THz region owe to diffraction effects due to the large wavelengths, as well as an increased free carrier and phonon absorption. In a microcavity, the population difference $N_1-N_2$ can be inferred precisely from the polariton splitting (the difference between the two polariton  frequencies) observed in the reflectivity experiment. Moreover, both methods require a careful determination of the linewidth of the ISB resonances.  

The impact of the dissipation effects on the spectra, such as free carrier absorption and ISB linewidth enlargement is visible in the data of Fig.\ref{Fig7}. For high temperatures the cavity resonances appear to have rather low quality factors, whereas the quality factors improve at low temperature, becoming comparable with the values measured for an undoped structure. This is clearly visible, e.g. in the case of the $s$=7.8$\mu$m cavity, where much narrower cavity resonance is recovered after the 2$\rightarrow$3 transition is depleted at very low temperature: the FWHM is 1.2 THz at 200 K and 0.79 THz at 7 K. This behavior confirms the comparison between the doped and undoped reference (Fig.\ref{Fig2}(a)). We attribute it mainly to the enlargement of the ISB resonances, that couple to the cavity mode, which is also visible in the multi-pass absorption spectra (Fig.\ref{Fig4}). Moreover, at high temperature the free carrier absorption in the heterostructure also increases,  as  part of the electronic population can be thermally excited in the high levels of the well and the 3D continuum of the heterostructure. 

\section{Data Exploitation}\label{PartIII}

To ascertain the above qualitative discussion we have modelled the experimental data for cavities with different dimensions at different temperatures. The data modelling is based on an effective medium approach, combined with numerical simulations of the optical response of the double metal structures. This approach allows us, on one hand, to consider the situation of several occupied subbands, and on the other, to take into account the finite linewidths of both the microcavity mode and the ISB resonances. The ohmic losses in the metallic layers are taken into account by considering a complex Drude constant for the metal \citep{Todorov_PRL_2009,PJouy_APL_2011} The effective medium constant can be obtained from the quantum Hamiltonian describing the light-matter interaction \citep{Todorov_PRB_2012}. In the following we start by recalling the quantum approach.  

\subsection{Plasma Hamiltonian and Effective medium approach}

The description of light-matter interaction using a Hamiltonian approach allows to take into account all the microscopic details of the electronic confinement. Moreover, the effective dielectric function of the quantum well media can be obtained from the characteristic polynomial of the quantum Hamiltonian. The quantum model allows therefore to obtain a dielectric function that is both directly applicable  for electromagnetic simulations and encapsulates the microscopic details of the system.  

To carry out this approach, we implement self-consistent Poisson-Schr\"{o}dinger coupled equations to calculate the potential changes associated with the field produced by the static charges. As a result, we obtain the envelope wave-functions $\phi_i(z)$ and energy levels $\hbar \omega_i$ produced by  the electronic confinement of the quantum wells, as wells as the Hartree correction. If the total sheet concentration $N_{tot}$ is known, we can then obtain the population of each subband $N_i$ through Fermi-Dirac statistic. We can then use this information in order to formulate the light-matter coupling Hamiltonian \citep{Todorov_PRB_2012}. The interaction of the transition $i\rightarrow j$ with light can be fully described introducing the function $\xi_{ij}(z)$, which represents the oscillating electronic current between levels $i$ and $j$:

\begin{equation}\label{2_xi}
\xi_{ij}(z) = \phi_i(z)\frac{\mathrm{d}\phi_j(z)}{\mathrm{d}z}-\phi_j(z)\frac{\mathrm{d}\phi_i(z)}{\mathrm{d}z}
\end{equation}

In the following, to simplify the notation, we use a single greek index $\alpha$ to indicate the transition $i\rightarrow j$. For instance, the frequencies of the ISB transitions are $\omega_\alpha= \omega_j-\omega_i$ . The current functions of Eq.\pref{2_xi} contain all the relevant quantities of the problem. Namely, the oscillator strength of the transition $\alpha$ is:

\begin{equation}\label{3_f}
f^o_\alpha = \frac{\hbar}{m^\ast \omega_\alpha}
\Big{(}\int \xi_{\alpha}(z){\mathrm{d}z}\Big{)}^2
\end{equation}

and the effective thickness $L_\alpha$ of the transition is:

\begin{equation}\label{4_L}
\frac{1}{L_\alpha} = \frac{\hbar}{m^\ast \omega_\alpha}
\int \xi_{\alpha}^2(z){\mathrm{d}z}
\end{equation}

Further, the effective length Eq.\pref{4_L} allows us to define the plasma frequency of the transition $\omega_{P\alpha}$:

\begin{equation}\label{5_omegaP}
\omega^2_{P\alpha} = \frac{e^2}{m^\ast \varepsilon \varepsilon_0}
\frac{\Delta N_\alpha}{L_\alpha}
\end{equation}

where $\Delta N_\alpha= N_i-N_j$ is the population difference (surface density) between levels $i$ and $j$. With the above definitions Eq.\pref{5_omegaP} coincides with the plasma frequency introduced by Ando, Folwer and Stern \citep{Ando_Fowler_Stern_1982}. Finally, we introduce the frequency of the intersubband plasmon $\widetilde{\omega}_\alpha$ which takes into account the depolarization effect:

\begin{equation}\label{6_omega tilde}
\widetilde{\omega}_\alpha = \sqrt{\omega^2_{\alpha}+\omega^2_{P\alpha}}
\end{equation}

These quantities permit to write the bosonized Hamiltonian describing the interaction of the intersubband plasmons with the cavity mode, as well as the coupling between the plasmons associated with different transitions \citep{Todorov_PRB_2012}:

\begin{eqnarray}\label{7_Hplasma}
\hat{H} =  \sum_{\alpha}\hbar
\widetilde{\omega}_\alpha p^\dagger_{\alpha}p_{\alpha} +\hbar
\omega_{\mathrm{cav}}(a^\dagger a + 1/2)\nonumber\\
+i\sum_{\alpha} \hbar \Omega_{\alpha}
(a^\dagger -a)(p^\dagger_{\alpha}+p_{\alpha})\nonumber\\
+\sum_{\alpha \neq \beta } \frac{1}{2} \hbar \Xi_{\alpha \beta}
(p^\dagger_{\alpha }+p_{\alpha})(p^\dagger_{\beta}+p_{\beta})
\end{eqnarray}

Here $p_{\alpha}$ and $p^\dagger_{\alpha}$ are bosonic operators describing the intersubband plasmons, $a$ and $a^\dagger$ are creation/annihilation operators of the cavity mode, and $\omega_{\mathrm{cav}}$ is the mode frequency. The strength of the light-matter coupling is provided by the coefficient $\Omega_{\alpha}$:

\begin{equation}\label{8_Om}
\Omega_{\alpha} = \frac{\omega_{P\alpha}}{2}\sqrt{\frac{\omega_{\mathrm{cav}}}{\widetilde{\omega}_\alpha} f^o_\alpha f^w_\alpha}
\end{equation}

where $f^w_\alpha = L_\alpha/L_{\mathrm{cav}}$ is the overlap between the microcavity and the intersubband plasmon. The strength of the inter-plasmon coupling is provided by the coefficient $\Xi_{\alpha \beta}$:

\begin{equation}\label{9_Xi}
\Xi_{\alpha \beta} = \frac{\omega_{P\alpha} \omega_{P\beta}}{2\sqrt{\widetilde{\omega}_\alpha \widetilde{\omega}_\beta}}\frac{\int \xi_\alpha (z)\xi_\beta (z) \mathrm{d}z}{\sqrt{\int \xi^2_\alpha (z)\mathrm{d}z \int \xi^2_\beta (z)\mathrm{d}z}}
\end{equation}

The Hamiltonian \pref{7_Hplasma} describes the coupling between set of quantum oscillators, one of which is the electromagnetic mode. The plasmonic oscillators are also coupled between each other through Coulomb dipole-dipole interactions, which are expressed in the last term of the Hamiltonian \pref{7_Hplasma}. As described in Appendix \ref{App_Multi_Band} the matter part of the Hamiltonian \pref{7_Hplasma} can be diagonalised numerically thus obtaining a set of independent plasmon modes each separately coupled with the light field:

\begin{eqnarray}\label{10_Hplasma}
\hat{H} =  \sum_{J}\hbar
W_J P^\dagger_J P_J +\hbar
\omega_{\mathrm{cav}}(a^\dagger a + 1/2)\nonumber\\
+i\sum_{J} \hbar \frac{R_J}{2}\sqrt{\frac{\omega_{\mathrm{cav}}}{W_J}}
(a^\dagger -a)(P^\dagger_J + P_J)
\end{eqnarray}

Here the $P^\dagger_J$ and $P_J$ are the new plasmonic operators with frequencies $W_J$. The coefficients $R_J$ play the role of effective plasma frequencies and characterize completely the coupling of the $J^{\mathrm{th}}$ mode to the electromagnetic field. As an illustration, let us consider the case of a single intersubband transition, $J = \alpha$. Then $R_J=R^0_\alpha$ becomes:

\begin{equation}\label{11_Rj}
R^0_\alpha = \omega_{P\alpha}\sqrt{f^o_\alpha f^w_\alpha}
\end{equation} 

This equation shows that the coefficient $R^0_\alpha$ contains altogether the information of the population differences $\Delta N_\alpha$ between the subband levels, through the plasma frequency Eq.\pref{5_omegaP}, as well as the overlap between the cavity modes and the intersubband polarization.

When several intersubband transitions are present, they interact through the last term of the Hamiltonian \pref{7_Hplasma}. After the Hamiltonian is diagonalized into its form \pref{10_Hplasma}, this interaction is also contained in the coefficients $R_J$. Moreover, owe to the plasmon-plasmon coupling, the coefficients $R_J$  are different from the ones that one would obtain for a single uncoupled plasmon (Eq. \pref{11_Rj}). This is precisely the phenomenon of oscillator strength transfer, as explained further.

As shown in details in Appendix \ref{App_Multi_Band}, the characteristic polynomial of the Hamiltonian \pref{10_Hplasma} allows to define the effective medium dielectric constant of the system:   

\begin{equation}\label{12_epEff}
\frac{\varepsilon}{\varepsilon_{\mathrm{eff}}(\omega)} = 1+\sum_J\frac{R^2_J}{\omega^2-W^2_J+i\omega\Gamma_J}
\end{equation}  
 
where we have introduced  the phenomenological linewidths, $\Gamma_J$, of the plasmon modes. This expression is similar to the plasmon pole approximation for 3D plasmons \citep{Book_Haug_Koch}. Note that, when rotating wave approximation is assumed, Eq.\pref{12_epEff} leads to a Lorenzian lineshape with full width at half-maximum equal to $\Gamma_J$. Once more, Eq.\pref{12_epEff} underlines the importance of the effective plasma frequencies $R_J$.  Indeed,  from the semi-classical point of view, $R^2_J$ is the weight of the plasmon pole $W_J$ of the inverse dielectric function, quantifying its oscillator strength. With this respect, the quantity $R^2_J$ can be considered as a generalization of the concept of oscillator strength in the case of ISB plasmons.

Note that, in the case of a multiple quantum well system as considered here, the Hamiltonian \pref{7_Hplasma} contains the couplings between plasmons from spatialy different quantum wells, as well as between plasmons from the same quantum well. However, in the case where we can neglect the overlap between wavefunctions from different quantum wells, the plasmon coupling coefficients \pref{9_Xi} vanish, and the remaining non-zero couplings are only those between plasmons from the same well. If the $N_{\mathrm{QW}}$ quantum wells are identical, we can consider a Hamiltonian of the form of Eq. \pref{10_Hplasma}, where the operators  $P^\dagger_J$ describe the normal plasmon modes of a single quantum well, but the plasma oscillator strengths are replaced by $R^2_J\times N_{\mathrm{QW}}$. This fact is briefly discussed in Appendix \ref{App_Multi_Band}.

To take into account the intrinsic anisotropy of the intersubband system and the in-plane free carrier absorption, the semiconductor slab is modeled by a diagonal dielectric tensor with components $\varepsilon_z = \varepsilon_{\mathrm{eff}}(\omega)$ and  $\varepsilon_x = \varepsilon_y = \varepsilon_{||}(\omega)$ \citep{Wendler_Kraft_PQW_1996}. The in-plane dielectric function $\varepsilon_{||}(\omega)$ describes the Drude absorption arising from the free movement of the carriers in the plane $(x,y)$\cite{Zanotto_APL_2010}. We have also considered the contribution of the LO phonons in the dielectric function, as in Ref.\citep{Todorov_JOSAA_2007}.

The knowledge of the effective medium dielectric constant expressed in Eq.\pref{12_epEff} gives the electromagnetic response of the ISB system, and therefore it is a central quantity for the interpretation of our data. In particular Eq.\pref{12_epEff} involves the quantities $W_J$, $\Gamma_J$ and $R_J$ describing the collective plasmon modes. As it is often the case we do not know all the exact details of the system and therefore we regard the quantities $W_J$, $\Gamma_J$ and $R_J$ as free parameters that can be determined by fitting the experimental spectra. Indeed, the positions of the intersubband peaks $W_J$ and their linewidths $\Gamma_J$ can be directly derived from the multi-pass absorption experiment (Fig.\ref{Fig4}). Moreover, the quantities $R_J$ can be deduced from the polariton splitting in measurement such as those presented in Fig.\ref{Fig7}.  

\begin{figure}
\includegraphics[scale=0.38]{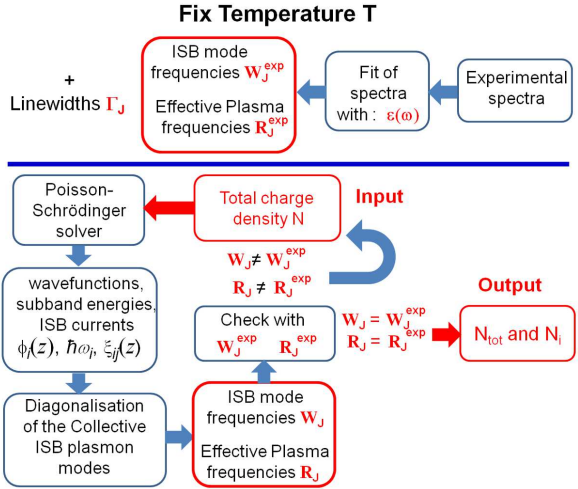}
\caption {Block diagram of the fitting procedure used to exploit the experimental data. Top: fitting the experimental spectra (reflectivity and absorption) with the electromagnetic model, based on the dielectric function of Eq.\pref{12_epEff} provides a phenomenological estimate for the quantities $R^{exp}_J$, $W^{exp}_J$ and $\Gamma_J$ for each temperature. Bottom: the quantities $R_J$, $W_J$ are computed from the microscopic quantum model, using the total number of electrons $N_{tot}$ in the well. For each temperature $T$ we retain the value of $N_{tot}$ which best reproduce the experimental values $R^{exp}_J$ and $W^{exp}_J$.} \label{Fig8}
\end{figure}

The values of $W_J$ and $R_J$ can also be calculated from the microscopic model described above, based on the Poisson-Schr{\"{o}}dinger modelling of the heterostructure and the diagonalization of the Hamiltonian of Eq. \pref{7_Hplasma}. In this approach, the critical input parameter is the number of ionized impurities that gives rise to the electronic charge (sheet) density $N_{tot}$. For each temperature $T$, the sheet density $N_{tot}$ is adjusted so that the calculated values of $W_J$, and $R_J$ coincide with the values obtained from the fit of the experimental spectra, $W^{exp}_J$, and $R^{exp}_J$. The scheme for relating the experimental spectra to the microscopic computation is summarized in Fig.\ref{Fig8}. Following this approach we can connect the results from the spectroscopic measurements with the microscopic information on the heterostructure as a function of the temperature $T$. In particular we can deduce the population of the different subbands $N_i$, as well as the bare intersubband energies $\hbar \omega_i$ and the corresponding single particle wavefunctions $\phi_i(z)$. 

We should bear in mind several underlying approximations. First, the quantum Hamiltonians of Eq.\pref{7_Hplasma} or Eq.\pref{10_Hplasma} describe a perfect electronic system, as they do not include any dissipative effects that may decrease or destroy the coherence of the electronic wave-functions, such as disorder or scattering from phonons or interface roughness. We approximate these effects by the inclusion of the phenomenological homogeneous broadening parameters $\Gamma_J$ in Eq.\pref{12_epEff}.  Moreover, the plasma Hamiltonians of Eq.\pref{7_Hplasma} and Eq.\pref{10_Hplasma} are valid in the limit of high electronic densities, where the exchange-correlation effect can be neglected.

\subsection{Modelling of the experimental data}

\begin{figure}
\includegraphics[scale=0.38]{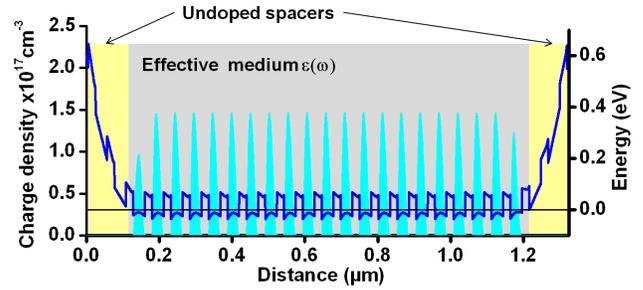}
\caption {Conduction band profile and charge density simulations of the system, taking into account the metal – semiconductor interfaces \cite{note_Greg_Snider}.} \label{Fig9}
\end{figure}

In our structure the crucial quantity for the evaluation of the light matter interaction is the total number of electron participating to process. In a metal-dielectric-metal cavity the Fermi level pins at the metal-GaAs interface \citep{Book_Sze} and therefore some of the quantum wells in the immediate vicinity of the metallic layers are depleted. In order to estimate the number of quantum wells that are still charged, $N_\mathrm{QW}$, we have used a freeware Poisson band-structure solver \cite{note_Greg_Snider}. The conduction band profile simulations are presented in Fig.\ref{Fig9}. The simulation shows that, below 150 K, 2 QWs are depleted at each interface. Both ends of the structure are therefore modelled as undoped spacers. The electronic distribution in the charged QWs is practically uniform, as can be seen from Fig.\ref{Fig9}. Indeed, we model this region with an effective dielectric constant of the form expressed in Eq.\pref{12_epEff}, which assumes an identical average doping for all the remaining 21 doped QWs. To take into account the spacer regions, according to Eq.\pref{11_Rj} the values for $R_J$ have been multiplied by $\sqrt{21/25}= 0.92$.

\begin{figure*}
\includegraphics[scale=0.58]{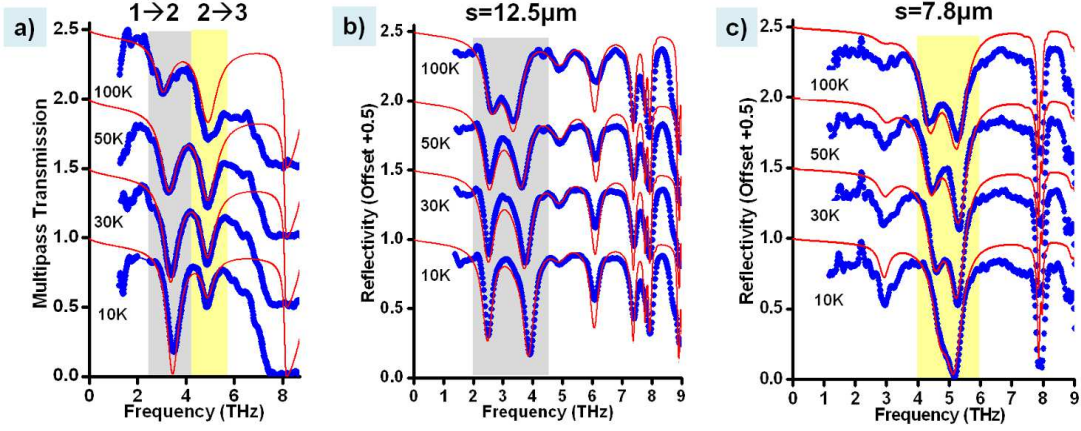}
\caption {Examples of electromagnetic simulations (solid curves) of the experimental spectra (dotted lines) for different temperatures. (a) Multipass transmission experiment. (b), (c) Reflectivity measurements at $\theta= 45^\circ$ for arrays of patches with dimensions $s=$12.5$\mu$m and $s=$7.8$\mu$m respectively.} \label{Fig10}
\end{figure*}

The total population $N_{tot}(T)$ and its redistribution on the different subbands as a function of the temperature $N_i(T)$ can be extracted by injecting it into the Hamiltonian \pref{10_Hplasma}, calculating $W_J$ and $R_J$ (for $J=1\rightarrow2$ and $J=2\rightarrow3$) and finally comparing these values with those obtained by the fit with the experimental spectra. Examples of the experimental fit are provided in Fig.\ref{Fig10}.

In Fig.\ref{Fig10}(a) the transmission experiment without cavity (dotted curve) is compared with a transfer matrix simulations based on the effective dielectric constant \pref{12_epEff} in order to extract the values of $W_J$ and $\Gamma_J$. These values are then inserted in the simulations of the reflectivity spectra presented in Figures \ref{Fig10}(b) and \ref{Fig10}(c), where we consider the particular case of cavities in resonance with the $1\rightarrow2$ and $2\rightarrow3$ ISB excitations respectively. From these simulations one can extract the effective plasma frequencies $R_{1\rightarrow2}$ and $R_{2\rightarrow3}$. Note that the reflectivity spectra from the cavities have intrinsically better quality than the multipass transmission data, especially at a high temperature. Indeed, in the multi-pass absorption light pass through a thick substrate (350$\mu$m) to probe a very thin active region (1.5$\mu$m). The substrate transparency is strongly affected by the increased optical phonon absorption as the temperature is raised. This is clearly visible in the data of Fig.\ref{Fig10}(a), where the transmission signal drops considerably around 7 THz. On the contrary, in the case of the cavity there is no substrate and the light field is crosses exclusively the QW region. This configuration is therefore more appropriate for testing ISB electronic absorption and avoiding spurious effect from the substrate. 

\begin{figure}
\includegraphics[scale=0.55]{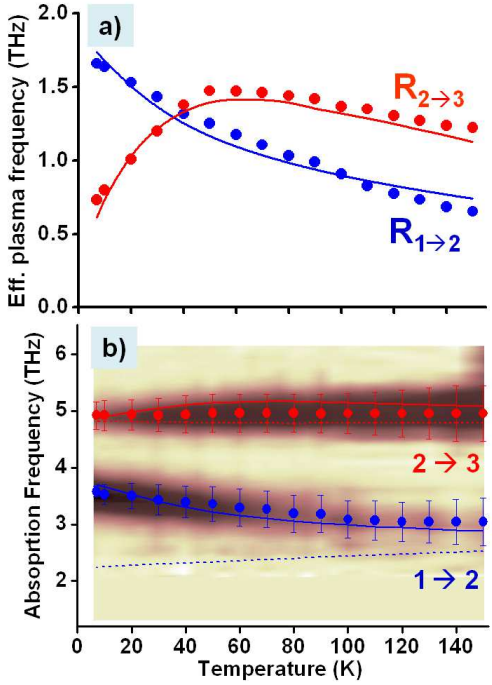}
\caption {(a) The effective plasma frequencies $R_{1\rightarrow2}$ and $R_{2\rightarrow3}$ for the two plasmon resonances, as extracted from experiments (dots) or computed from the quantum approach (solid lines). (b) Measured plasmon resonances $W_{1\rightarrow2}$, $W_{2\rightarrow3}$ and their line widths $\Gamma_{1\rightarrow2}$ , $\Gamma_{2\rightarrow3}$ (dots and error bars), as compared to the results from the quantum approach. In dashed lines: the subband separations $\omega_{1\rightarrow2}$ and $\omega_{2\rightarrow3}$ of the bare electronic states.} \label{Fig11}
\end{figure}

\begin{figure}
\includegraphics[scale=0.53]{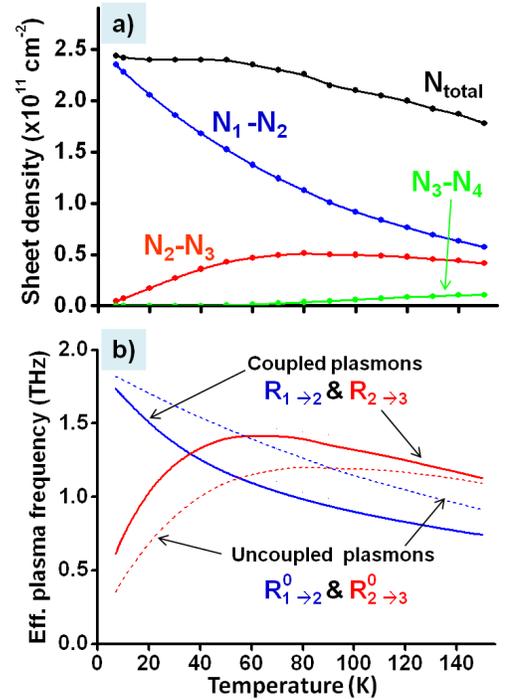}
\caption {(a) Population differences between different subband levels, as wells as total electronic sheet density as a function of the temperature, as computed from the quantum model. (b) Comparison between the effective plasma frequencies of the two resonances in the case without coupling (dashed lines) and when the plasmon-plasmon coupling is taken into account (solid lines).} \label{Fig12}
\end{figure}

The parameters $W_J$, and $R_J$ obtained from this procedure have been plotted as a function of the temperature in Fig.\ref{Fig11}(a) and Fig.\ref{Fig11}(b) (full circles). In Fig.\ref{Fig11}(b) the full widths at half maximum of the absorption peaks, $\Gamma_J$, are also indicated as errors bars. The continuous lines are the results of our simulation using a Poisson-Schr\"{o}dinger solver described in Appendix \ref{App_PS_solver} together with the diagonalization of the plasmon Hamiltonian \pref{7_Hplasma}. As we have already mentioned, we have left, for each temperature $T$, the total sheet electronic density per well as a free fitting parameter. In Fig.\ref{Fig12}(a) we present $N_{tot}$ as a function of the temperature as well as the population differences, $(N_i-N_j)(T)$ between the occupied subbands. For the fit we have used the value of 31.8nm for the quantum well thickness (instead if the nominal value 32nm), which gives us a good fit over the whole temperature range. According to these results the number of electrons $N_{tot}(T)$ captured by the quantum wells is constant below 60 K, and then progressively decreases as the temperature is raised. This behaviour can be explained by assuming that some of the electrons are recaptured in the delta doped donor region. 

In Fig.\ref{Fig11}(b) we also report the frequencies $\omega_{1\rightarrow2}$ and $\omega_{2\rightarrow3}$ (dashed lines) calculated as the difference between the single particle energy states. Due to the Hartree potential in a modulation doped structure $\omega_{1\rightarrow2}$ decreases as the electronic population is increased. Therefore, as the temperature is lowered between 7 K and 60 K, the energy spacing $\hbar \omega_{1\rightarrow2}$ is reduced and the subband 2 becomes populated. Note also that the collective plasma frequency $W_{1\rightarrow2}$, that fits very well the experimental data, has exactly the opposite behavior with respect to $\omega_{1\rightarrow2}$. In the case of the $2\rightarrow3$ transition the depolarization effect is much smaller and the collective mode frequency $W_{2\rightarrow3}$ is close to the single particle frequency $\omega_{2\rightarrow3}$. The observed slight depolarization shift for this transition is even lower than that predicted by the quantum Hamiltonian \pref{7_Hplasma} that describes a perfectly coherent electronic system ($\Gamma_j = 0$). The small discrepancy arises from disorder effects \cite{SLuin_2001}. As a matter of fact it can be noticed that $W_{2\rightarrow3}-\omega_{2\rightarrow3} < \Gamma_{2\rightarrow3}$, which implies that the correlations introduced by the depolarization effects are not enough to overcome the disorder, and the system stays closer to its single particle energy. In this case the effects of the inhomogeneous broadening become important, and models like those exposed in Ref. \cite{Metzner_Dohler_1999} are more pertinent to describe the system.

The microscopic model used to fit the experimental data permits also to evidence the interaction between ISB plasmons from different transitions which leads to transfer of the oscillation strength between the first and excited ISB transitions \citep{Warburton_1998}. Let us recall how this phenomenon is contained in the Hamiltonians \pref{7_Hplasma} and \pref{10_Hplasma}. The Hamiltonian \pref{7_Hplasma} describes a set of collective modes at frequencies $\widetilde{\omega}_\alpha$. Their plasma frequencies  $\omega_{P\alpha}$ (Eq.\pref{5_omegaP}) not only quantify their coupling strength with the cavity mode, but also describe their interaction through the dipole-dipole Coulomb coupling, Eq.\pref{9_Xi}. By diagonalizing the matter part of the Hamiltonian \pref{7_Hplasma}, we obtain the Hamiltonian \pref{10_Hplasma} which contains the new independent normal collective modes at frequencies $W_J$.  Therefore the light-matter interaction occurs between plasmons originating from individual ISB plasmons coupled by Coulomb interaction. A side effect of the coupling is that, when two ISB excitations are present, as in our system, the first excitation transfers part of its oscillator strength to the second one.

For our system this effect is not visible in the excitation frequencies, as the normal mode frequencies $W_{1\rightarrow2}$ and $W_{2\rightarrow3}$ remain close to the original collective mode frequencies $\widetilde{\omega}_{1\rightarrow2}$ and $\widetilde{\omega}_{2\rightarrow3}$ (not shown). However, the transfer of oscillator becomes apparent by considering the plasma energies. This is illustrated in Fig.\ref{Fig12}(b), where we compare $R_{1\rightarrow2}$ and $R_{2\rightarrow3}$ (continuous line) with the values $R^0_{1\rightarrow2}$ and $R^0_{2\rightarrow3}$ that would be obtained for two completely uncoupled plasmons (dashed line). The later can be obtained analytically from Eq.\pref{11_Rj}. We observe that always $R_{2\rightarrow3} > R^0_{2\rightarrow3}$ and $R_{1\rightarrow2} < R^0_{1\rightarrow2}$. Notice that at low temperature $R_{2\rightarrow3}$ is almost twice $R^0_{2\rightarrow3}$. This means that the effective oscillator strength of the plasmon resonance is doubled when we consider the Coulomb dipole-dipole interaction. This explains why $2\rightarrow3$ absorption peak in the spectra of Fig.\ref{Fig10}(a) is quite high in spite of the relatively low population on the subband 2.  

\subsection{Photonic dispersion and finite linewidth effects}

\begin{figure}
\includegraphics[scale=0.40]{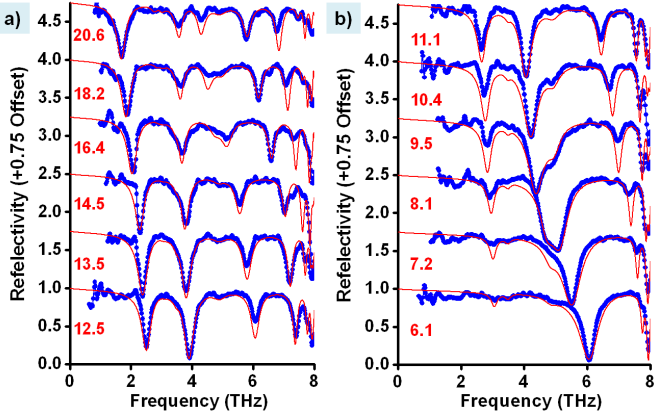}
\caption {(a) and (b) Experimental reflectivity spectra at $\theta= 10^\circ$ (dotted curves) and simulations (solid lines) for all measured cavities, at low temperature $T=$7 K. The number next to each spectra indicates the size of the patch $s$.} \label{Fig13}
\end{figure}

To proof the consistency of the values of $W_J$ and $R_J$ extracted from the fits in Fig.\ref{Fig10} we have simulated the spectra for all the measured cavities, at four different temperatures: 7 K, 30 K, 50 K and 100 K. The cavities used in the experiments had a patch width ranging from 6 to 21$\mu$m. This permits to explore a frequency range from 1.8 THz to 6 THz for the fundamental mode. The comparison between the experimental spectra (dotted lines) and the effective medium electromagnetic simulations (continuous lines) are presented in Fig.\ref{Fig13} at $T=$7 K. For simplicity, we consider only measurements with an incident angle of $\theta= 10^\circ$, where only the odd resonances are excited. The agreement between the experiments and electromagnetic simulations is excellent. The simulations reproduce correctly both the positions of the different resonances and the evolution of their reflectivity dips.

\begin{figure}
\includegraphics[scale=0.42]{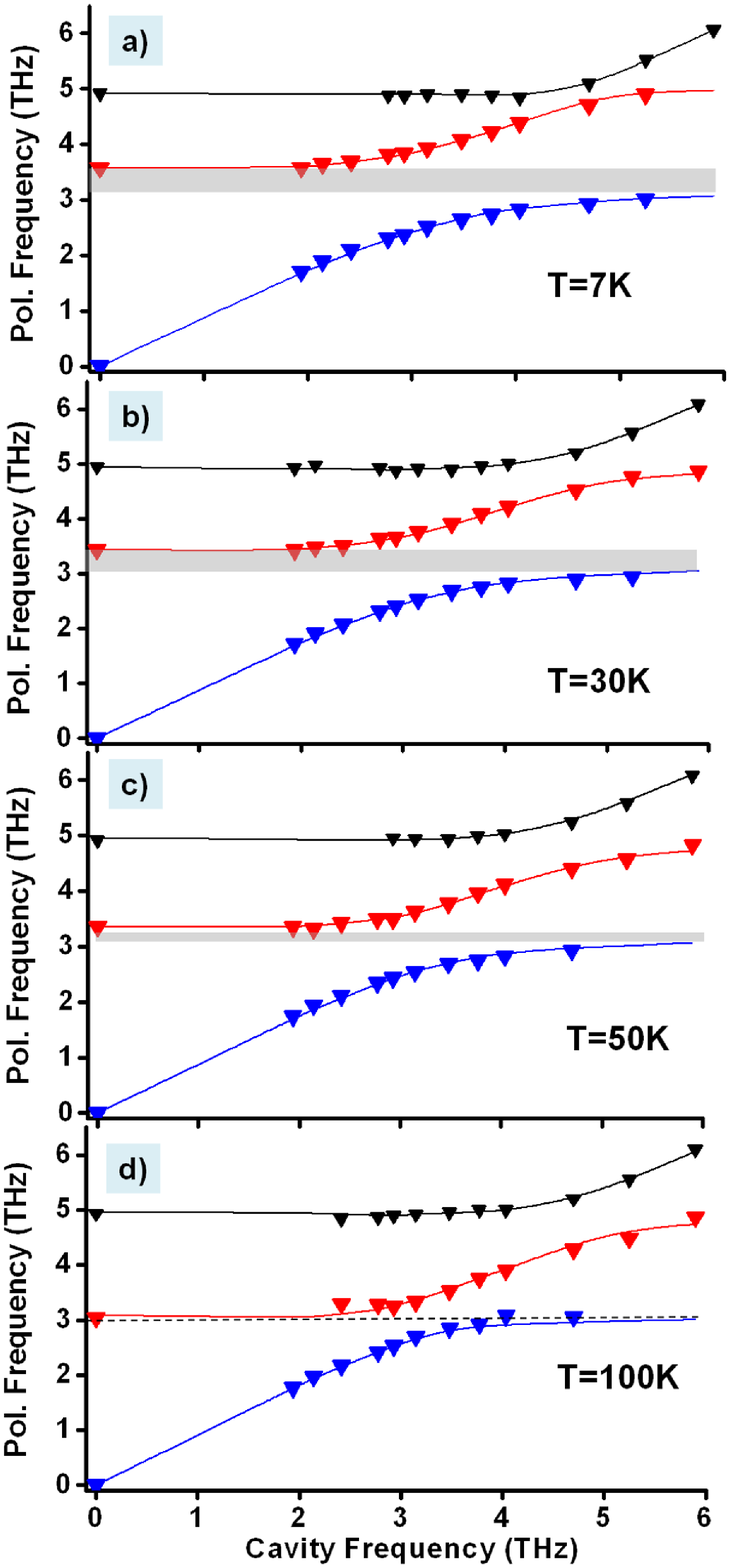}
\caption {Dispersion curves for the coupling between the two ISB plasmons with the fundamental cavity mode. The solid lines are electromagnetic simulations, and the triangles are  measured polariton frequencies.  (a) $T=$7 K, (b) $T=$30 K, (c) $T=$50 K, (d) $T=$100 K. The grey area indicates the polariton gap for the $1\rightarrow2$ ISB plasmon.} \label{Fig14}
\end{figure}

To summarize the comparison between measurements and experiments in Fig.\ref{Fig14} we have plotted the polariton dispersion curves for the four different temperatures above mentioned. Again, very good agreement is obtained between the experiment (triangles) and the effective medium model based upon the dielectric constant Eq.\pref{12_epEff} (continuous lines) for every temperature. These dispersion curves indicate clearly how the splitting of the high energy excitation $2\rightarrow3$ increases while that of the $1\rightarrow2$ excitation decreases as the temperature is raised. In accordance the polariton gap associated with the $1\rightarrow2$ excitations progressively decreases with the temperature and at $T =$100 K no gap can be seen. The "temperature knob" therefore allows us to tune the system from the ultra-strong coupling to the "ordinary" strong-coupling regime \citep{PJouy_2011}.

The study of the polaritonic dispersion brings insights on the impact of the dissipation (linewidth enlargement) on the measured splitting. In a system with loss the observable Rabi splitting is in general smaller than the actual strength of the light-matter interaction \citep{Savona1995733}. For an ideal lossless system, the polariton dispersion can be obtained analytically from the eigenvalue equation of the quantum Hamiltonian \citep{Todorov_PRB_2012}. For our particular system, the dispersion equation for two plasmon modes coupled with a lossless $\mathrm{TM}_0$ mode is:

\begin{equation}\label{13_Rdisp}
1+\frac{R^2_{1\rightarrow 2}}{\omega^2-W^2_{1\rightarrow  2}}+\frac{R^2_{2\rightarrow 3}}{\omega^2-W^2_{2\rightarrow 3}} = \frac{\omega^2}{\omega^2_\mathrm{cav}}
\end{equation}
 
\begin{figure}
\includegraphics[scale=0.35]{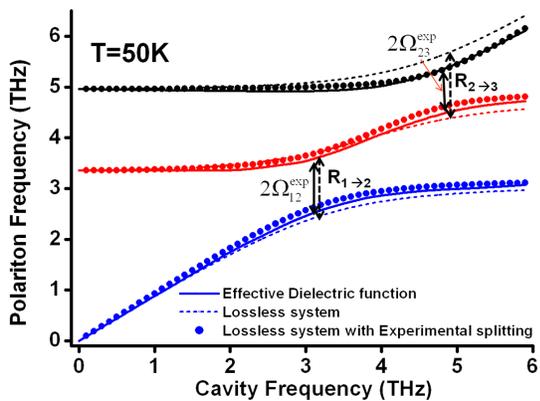}
\caption {Dispersion curves computed from the different models mentioned in the text.} \label{Fig15}
\end{figure}

The real roots of Eq.\pref{13_Rdisp} are plotted in a dashed-line in Fig.\ref{Fig15}, for $T=$50 K, where both ISB excitations are clearly present. The results from the dielectric constant model, that fits exactly the measured data, are indicated as a solid-line. Not surprisingly the predicted Rabi splittings for the ideal system described by Eq.\pref{13_Rdisp} are greater than those obtained experimentally. This is the results of the reduced coherence in the real system due to the loss introduced by the metallic layers of the cavity, the finite width of the ISB resonance, as well as the decreased transparency of the semiconductor close to the restrahlen band. 

For an infinitely coherent system, according to Eq.\pref{13_Rdisp} the polariton splittings $2\Omega^{exp}_{12}$ and $2\Omega^{exp}_{23}$ would be exactly $R_{1\rightarrow2}$ and $R_{2\rightarrow3}$. Interestingly, a very good fit will be obtained also by replacing in Eq.\pref{13_Rdisp} $R_{1\rightarrow2}$ and $R_{2\rightarrow3}$ with the experimental values for the splittings ($2\Omega^{exp}_{12}$ and $2\Omega^{exp}_{23}$) as it is shown in the dotted-curve of Fig.\ref{Fig15}. This implicates that the polaritonic dispersion can be very well reproduced from the perfect quantum model, as long as the measured values of the splittings $2\Omega^{exp}_{i}$  are used for the light-matter coupling constants in the Hamiltonian. Moreover, it justifies the approach of several experimental groups that fit the measured dispersion curves with fully coherent equations, like Eq.\pref{13_Rdisp}, by using the experimentally derived Rabi frequency at resonance \cite{Todorov_PRL_2009,Anappara_2009a,PJouy_2011,Geiser_2012}.  

\begin{table}
\caption{\label{Table_I}
$1\rightarrow2$ transition}
\begin{ruledtabular}
\begin{tabular}{cccccc}
$T$ (K) &$R_{1\rightarrow2}$ \footnotemark[1] & $2\Omega^{exp}_{12}$ \footnotemark[2] & $\Gamma_{1\rightarrow2}$ \footnotemark[1] & $\Gamma_{\mathrm{cav}}$ \footnotemark[3] 
 & $C_{1\rightarrow2}$ \\
\hline
7 & 1.64 & 1.40 & 0.20  & 0.55 & 12.77 \\
30 & 1.41 & 1.20 & 0.53  & 0.55 & 3.4 \\
50 & 1.23 & 1.04 & 0.60  & 0.55 & 2.3 \\
100 & 0.90 & 0.70 & 0.65  & 0.55 & 1.1
\end{tabular}
\end{ruledtabular}
\footnotetext[1]{In THz, from the effective medium model.}
\footnotetext[2]{In THz, from the dispersion curves.}
\footnotetext[3]{In THz, simulation of a $s=$12.5$\mu$m cavity without carriers.}
\end{table}

\begin{table}
\caption{\label{Table_II}
$2\rightarrow3$ transition}
\begin{ruledtabular}
\begin{tabular}{cccccc}
$T$ (K) &$R_{2\rightarrow3}$ \footnotemark[1] & $2\Omega^{exp}_{23}$ \footnotemark[2] & $\Gamma_{2\rightarrow3}$ \footnotemark[1] & $\Gamma_{\mathrm{cav}}$ \footnotemark[3] 
 & $C_{2\rightarrow3}$ \\
\hline
7 & 0.72 & 0.24 & 0.48  & 0.77 & 0.7 \\
30 & 1.18 & 0.61 & 0.58  & 0.77 & 1.6 \\
50 & 1.46 & 0.80 & 0.68  & 0.77 & 2.0 \\
100 & 1.35 & 0.82 & 0.70  & 0.77 & 1.7
\end{tabular}
\end{ruledtabular}
\footnotetext[1]{In THz, from the effective medium model.}
\footnotetext[2]{In THz, from the dispersion curves.}
\footnotetext[3]{In THz, simulation of a $s=$7.8$\mu$m cavity without carriers.}
\end{table}

The comparison between the ideal system and our real system is summarized in Table \ref{Table_I} and Table \ref{Table_II}, for the $1\rightarrow2$ transition and $2\rightarrow3$ transition, respectively. A sizeable discrepancy can be measured between the modelled coupling constant $R_{2\rightarrow3}$ and the observable splitting, $2\Omega^{exp}_{23}$, particularly when the linewidth is an important fraction of these quantities. To quantify this effect we define cooperativity parameter, as in atomic physiscs, that sets the onset for the strong coupling in our system. In atomic physics, where an atom is interacting with a microcavity mode, the cooperativity parameter is defined as the ratio between the rate of photon scattering into the microcavity mode divided by the rate of the free-space scattering (radiative enlargement of the atomic transition) \cite{Wickenbrock_2011, McKeever_2003}. This definition can be adapted to our system by taking into account the fact that the dominant contribution to linewidth enlargement is not of radiative origin. Indeed, the decay rate of the ISB plasmon is $\Gamma_J/2$ and is determined mainly by non-radiative mechanisms, such as phonon relaxation and interface scattering. We can define thus the cooperativity parameter $C_J$ as:

\begin{equation}\label{14_CJ}
C_J = \frac{R^2_J}{2\Gamma_{\mathrm{cav}}\Gamma_J}
\end{equation}  

where  $R^2_J/4\Gamma_{\mathrm{cav}}$ is the rate of photon emission in the cavity mode as provided by Fermi Golden rule \cite{Ciuti_Carusotto_2006}, since the coupling constant of the Hamiltonian of Eq.\pref{10_Hplasma} is $R_J/2$ and $\Gamma_{\mathrm{cav}}$ is the full-width at half maximum of the cavity mode. With this definition, the criterion for the intersubband system to be in the strong coupling regime is $C_J > 1$.

The values of $C_J$ are reported in the rightmost column of the Tables \ref{Table_I} and \ref{Table_II}, and show that this definition is in accordance with our data. In particular we can observe that the coupling between the $2\rightarrow3$ transition and the fundamental mode has in general a lower cooperativity and therefore a much higher discrepancy between the values of the coupling constant $R_J$ and the measured splittings $2\Omega^{exp}_{23}$ is observed. This is due to the dissipations effect of the $2\rightarrow3$ transition being more important than those of the $1\rightarrow 2$ transition, probably due to the proximity of the GaAs phonon absorption band. According to the criterion $C_J > 1$, the $2\rightarrow3$ transition is not in strong coupling at $T=$7 K as it can be also seen in Fig.\ref{Fig10}(c) and Fig.\ref{Fig14}(a).  More generally, the importance of the dissipation effects on the onset of the strong coupling could explain the discrepancies between the light-matter coupling estimated from simple analytical formulas like Eq.\pref{1_Splitting} and experimental observations, especially in a case of large ISB linewidths or of metallic cavities with high losses as those reported in Ref.\cite{Geiser_2012}.

\section{Conclusion}

In conclusion, we have presented a study of intersubband systems strongly coupled with 0D micro-resonators. Our analysis permits to relate the polariton spectroscopic features to microscopic details of the heterostructure, such as the thermal distribution of electrons on the different subbands and the coupling between different collective oscillations in the quantum well. We interpret our experimental results using an effective medium approach, which connects the quantum simulation of the heterostructure with the electromagnetic modelling of the microresontor containing the quantum well slab. We also discuss some of the important aspects of the experimental system, such as finite linewidths or charge transfer occurring at metal-semiconductor interfaces. Such modelling can be further refined by including, for instance, inhomogeneous broadening effects owed to disorder like in the numerical approach proposed in Ref.\cite{Metzner_Dohler_1999}. Since we consider samples with high electronic densities, we expect that other fundamental effects, like exchange-correlation and excitonic like shift \cite{Manasreh_1991} to be negligible.

We therefore believe that the approach described here, which is a combination of microscopic self-consistent modelling of the heterostructure and electromagnetic simulations through the effective medium constant (Eq.\pref{12_epEff}) could be applied as a designer tool for ISB light-matter coupled quantum devices. 

We gratefully acknowledge support from the French National Research Agency in the frame of its Nanotechnology and Nanosystems program P2N, Project No. ANR-09-NANO-007. We acknowledge financial support from the ERC grant "ADEQUATE", and from the Austrian Science Fund (FWF). 

\appendix

\section{Periodic Poisson-Schr\"{o}dinger solver}\label{App_PS_solver}

In a typical self-consistent Poisson-Schr{\"{o}}dinger problem one determines the Hartree correction of the potential $V_H(z)$, that includes the contribution of the static charges through the Poisson equation \citep{Ando_Fowler_Stern_1982}. This contribution is added to the heterostructure potential, and the Schr{\"{o}}dinger equation is solved numerically, typically trough the shooting method. 

As seen from Fig.\ref{Fig9}, the doped quantum wells create a periodic potential. Hence the Hartee correction $V_H(z)$ to the heterostructure potential, that depends on the wavefunctions is also periodic. However, such self-consistent Poisson-Schr{\"{o}}dinger problem is inconvenient to treat with the shooting method, since the energy levels of the quantum wells are heavily degenerated. We therefore treat only a single period with extension $d$ (such as $V_H(z+d)= V_H(z)$). The most general solution for the Poisson equation for a 2D charge density $\rho(z)$ is then \citep{Ando_Fowler_Stern_1982}:

\begin{equation}\label{A1_1}
V_H(z) = \frac{e^2}{\varepsilon \varepsilon_0}\int_0^z (z-z')\rho(z')\mathrm{d}z' + Az+B
\end{equation}  

Here $\rho(z)$ contains both the contribution from the wavefunctions and the donor impurities. We choose $B=0$ and we require that the charge density is also periodic: $\rho(z+d)= \rho(z)$.  Moreover, we require that the starting point $z=0$ of the period is chosen so that the period is globally neutral:

\begin{equation}\label{A1_2}
\int_0^d \rho(z')\mathrm{d}z' =0
\end{equation} 

Hence the constant $A$ is uniquely determined and Eq.\pref{A1_1} becomes:

\begin{equation}\label{A1_3}
V_H(z) = \frac{e^2}{\varepsilon \varepsilon_0}
\Big{[}
\int_0^z (z-z')\rho(z')\mathrm{d}z' +\frac{z}{d}\int_0^d z'\rho(z')\mathrm{d}z'
\Big{]}
\end{equation} 

It is easy to verify with this formula that both the potential $V_H(z)$ and its derivative $\mathrm{d}V_H(z)/\mathrm{d}z$ are $d$-periodic. Hence Eq.\pref{A1_3} describes the unique solution of the Poisson equation on a globally neutral cylindrical space with a circumference $d$.

\section{Multiband Hamiltonian}\label{App_Multi_Band}

The diagonalization of the Hamiltonian \pref{7_Hplasma} is performed in two steps. First, we diagonalize the matter part of the Hamiltonian which includes the inter-plasmon interactions. The result is a set of new independent normal collective plasmon modes, which are coherent superpositions of plasmons from different transitions. We then explicit the interaction with the normal modes with the light mode. The new coupling constants of the light-matter interaction then describe the collective oscillator strength of the normal modes.   

We therefore start with the matter Hamiltonian:

\begin{eqnarray}\label{A2_1}
\hat{H}_{mat} =  \sum_{\alpha}\hbar
\widetilde{\omega}_\alpha p^\dagger_{\alpha}p_{\alpha} \nonumber \\
+ \sum_{\alpha \neq \beta } \frac{1}{2}\hbar \Xi_{\alpha \beta}
(p^\dagger_{\alpha }+p_{\alpha})(p^\dagger_{\beta}+p_{\beta})
\end{eqnarray}

If  there are $N$ optically active plasmons in the system, then, after the diagolization is performed, there will be $N$ independent normal modes.  The  bosonic operators $P_J$ of the normal modes are linear combination of the initial ones: 

\begin{equation}\label{A2_2}
P_J = \sum_\alpha (Y_{J\alpha} p_\alpha +T_{J\alpha}p^\dagger_\alpha)
\end{equation}  

The latter are eigenmodes of the Hamiltonian \pref{A2_1} with eigenfrequencies $W_J$ only if:

\begin{equation}\label{A2_3}
[P_J, \hat{H}_{mat}] =\hbar W_J P_J
\end{equation}

Using the symmetry property $\Xi_{\alpha \beta}=\Xi_{\beta \alpha}$ and the relation:

\begin{equation}\label{A2_4}
[p_\alpha, \hat{H}_{mat}] =\hbar \widetilde{\omega}_\alpha p_\alpha + \sum_{\alpha \neq \beta } \hbar \Xi_{\alpha \beta}(p^\dagger_{\beta}+p_{\beta})
\end{equation}

as well as its hermitian conjugate, Eq.\pref{A2_3} leads to the following eigenvalue problem:    

\begin{equation}\label{A2_5}
\mathbf{M} \mathbf{V}_J = W_J \mathbf{V}_J
\end{equation}     

Here $\mathbf{M}$ is the $2N\times 2N$ Hopfield-Bogoliubov matrix of the problem, and and $\mathbf{V}_J$ is the $2N$ rank eigenvector which corresponds to the eigenvalue $W_J$:

\begin{equation}\label{A2_6}
\mathbf{M}=
\left( \begin{array}{cccc}
\mathbf{I}_1 & \mathbf{C}_{12} & \cdots & \mathbf{C}_{1N} \\
\mathbf{C}_{12} & \mathbf{I}_2 & \cdots & \mathbf{C}_{2N} \\
\vdots & \vdots & \ddots & \vdots \\
\mathbf{C}_{1N} & \mathbf{C}_{2N} & \cdots & \mathbf{I}_{N} \\
\end{array}\right )
\end{equation}

\begin{equation}\label{A2_6a}
\mathbf{I}_\alpha =
\left[ \begin{array}{cc}
-\widetilde{\omega}_\alpha & 0 \\
0 & \widetilde{\omega}_\alpha
\end{array}\right ] \phantom{QQ}
\mathbf{C}_{\alpha \beta} =
\left[ \begin{array}{cc}
-\Xi_{\alpha \beta} & -\Xi_{\alpha \beta} \\
\Xi_{\alpha \beta} & \Xi_{\alpha \beta}
\end{array}\right ]
\end{equation}

\begin{equation}\label{A2_6b}
\mathbf{V}_J = ^T(Y_{J1},T_{J1}, \cdots Y_{JN},T_{JN})
\end{equation} 

Since the matrix $\mathbf{M}$ real and block-symmetric, it allows reel eigenvalues with reel eigenvectors. Moreover, if $W_J$ is an eigenvalue, then $-W_J$  is also eigenvalue of $\mathbf{M}$. Therefore the matrix $\mathbf{M}$ has $N$ distinct positive eigenvalues  $W_J$ which are the frequencies of the normal plasmon modes. The latter can be easily inferred from the numerical solution of the linear problem described by Eq.\pref{A2_5}.

Let us now explicit the coupling of the normal modes with the light mode. Since the eigenvectors $\mathbf{V}_J$ are real, then it is easy to show that we have:

\begin{equation}\label{A2_7}
P_J + P_J^\dagger = \sum_\alpha (Y_{J\alpha}+T_{J\alpha})(p_\alpha+ p^\dagger_\alpha)
\end{equation}  

This relation, seen as a matrix-vector product, allows to determine $p_\alpha+ p^\dagger_\alpha$  as a function of  $P_J + P_J^\dagger$. Let us define the $N\times N$ matrix inverse:                                    

\begin{equation}\label{A2_8}
X_{\alpha J} = (Y_{J\alpha}+T_{J\alpha})^{-1}
\end{equation}  

This leads to the light-matter coupling Hamiltonian of Eq.\pref{10_Hplasma}, where the effective plasma frequencies $R_J$ of the normal modes are defined as:

\begin{equation}\label{A2_9}
R_J = \sum_\alpha \omega_{P\alpha} X_{\alpha J} \sqrt{\frac{W_J}{\widetilde{\omega}_\alpha}f_\alpha^o f_\alpha^w}
\end{equation}  

At this point, the Hamiltonian \pref{10_Hplasma} describes $N$ independent quantum oscillators coupled with a single cavity mode. Its Hopfield matrix is very similar to 
Eq. \pref{A2_6}, except that the coupling terms are non zero only on the first two columns and first two rows. The characteristic polynomial of this problem can then be computed explicitly, and the result is: 

\begin{equation}\label{A2_10}
\prod_J(\omega^2-W_J^2)\Big{\{ }\omega^2-\omega_{\mathrm{cav}}^2 -\sum_J \frac{R^2_J \omega_{\mathrm{cav}}^2}{\omega^2-W_J^2}  \Big{ \} }
\end{equation}

Therefore the equation providing the light-coupled polariton states is:

\begin{equation}\label{A2_11}
\omega^2-\omega_{\mathrm{cav}}^2 -\sum_J \frac{R^2_J \omega_{\mathrm{cav}}^2}{\omega^2-W_J^2}= 0
\end{equation}

In the case where there are $N_\mathrm{QW}$ initial plasmons are uncoupled, but oscillate at identical frequencies, the above formalism predicts that there is one bright excitation, with a collective oscillator strength that is $\sqrt{N_\mathrm{QW}}$ higher. This is easily seen from the above equation, by setting $W_J = \mathrm{const}$.    

The effective dielectric constant of the system is then defined through the relation  $\varepsilon_{\mathrm{eff}}(\omega)\omega^2 = \varepsilon \omega_{\mathrm{cav}}^2$:

\begin{equation}\label{A2_12}
\frac{\varepsilon}{\varepsilon_{\mathrm{eff}}(\omega)} = 1+\sum_J \frac{R^2_J}{\omega^2-W^2_J}
\end{equation} 

The treatment of the problem is so far energy conserving. When introducing Markovian  relaxation mechanism with a rate $\Gamma_J/2$ and considering a basis of coherent (semiclassical) states the above equation is transformed into Eq.\pref{12_epEff} \citep{Dutra_Furuya_1998}.

Let us discuss briefly the case of a heterostructure consisiting of $N_\mathrm{QW}$ identical quantum wells. For simplicity, let us consider only two plasmons from each quantum well, coupled by a constant $\Xi_{a b}$, and we consider that the plasmons from different wells are uncoupled, owe to a vanishing overlap of the wavefunctions between different wells. In this case the matter Hamiltonian writes:

\begin{eqnarray}\label{A2_15}
\hat{H}_{mat} =  \sum_{i=1}^{N_\mathrm{QW}}\hbar
\widetilde{\omega}_a p^\dagger_{a, i}p_{a, i} 
+ \sum_{i=1}^{N_\mathrm{QW}}\hbar
\widetilde{\omega}_b p^\dagger_{b, i}p_{b, i}
\nonumber \\
+ \frac{1}{2}\hbar \Xi_{a b} \sum_{i=1}^{N_\mathrm{QW}} 
(p^\dagger_{a,i}+p_{a,i})(p^\dagger_{b,i}+p_{b,i})
\end{eqnarray} 

The light matter-part interaction Hamiltonian is respectivelly

\begin{eqnarray}\label{A2_16}
\hat{H}_{int} =  i\hbar \Omega_{a} \sum_{i=1}^{N_\mathrm{QW}}  
(a^\dagger -a)(p^\dagger_{a,i}+p_{a,i})\nonumber\\
+i\hbar \Omega_{b} \sum_{i=1}^{N_\mathrm{QW}}  
(a^\dagger -a)(p^\dagger_{b,i}+p_{b,i})
\end{eqnarray} 

To simplifiy these Hamiltonians, we introduce the following new bosonic operators \citep{Ciuti_PhysRevB_2005}:

\begin{eqnarray}\label{A2_17}
\pi^\dagger_{a,b} = \frac{1}{\sqrt{N_\mathrm{QW}}}\sum_{i=1}^{N_\mathrm{QW}}p^\dagger_{(a,b),i}
\end{eqnarray} 

Here the normalization factor $1/\sqrt{N_\mathrm{QW}}$ arises from the requirement that $[\pi_{a,b},\pi^\dagger_{a,b}]=1$. Let us then compute the commutator of $\pi^\dagger_{a}$ with the Hamiltonian \pref{A2_15}:

\begin{eqnarray}\label{A2_18}
[\hat{H}_{mat},\pi^\dagger_{a}] = \hbar \widetilde{\omega}_a 
\frac{1}{\sqrt{N_\mathrm{QW}}}\sum_{i=1}^{N_\mathrm{QW}} p^\dagger_{a, i} \nonumber\\ +
\frac{1}{2}\hbar \Xi_{a b} \frac{1}{\sqrt{N_\mathrm{QW}}} \sum_{i=1}^{N_\mathrm{QW}} 
(p^\dagger_{b,i}+p_{b,i}) = \nonumber\\
\hbar \widetilde{\omega}_a  \pi^\dagger_a + \frac{1}{2}\hbar \Xi_{a b} (\pi_b+\pi^\dagger_b)
\end{eqnarray}

Similar commutation relation is valid for $\pi^\dagger_b$. These commutation relations are identical if we replace Eq. \pref{A2_15} with the effective Hamiltonian:

\begin{eqnarray}\label{A2_19}
\hat{H}_{mat} = \hbar \widetilde{\omega}_a \pi^\dagger_a \pi_a 
+ \hbar \widetilde{\omega}_b \pi^\dagger_b \pi_b \nonumber\\ +
 \frac{1}{2}\hbar \Xi_{a b} 
(\pi^\dagger_a+\pi_a)(\pi^\dagger_b+\pi_b)
\end{eqnarray}  

Expressing the interaction Hamiltonian with the new operators we obtain:

\begin{eqnarray}\label{A2_20}
\hat{H}_{int} =  i\hbar \Omega_{a} \sqrt{N_\mathrm{QW}}  
(a^\dagger -a)(\pi^\dagger_a+\pi_a)\nonumber\\
+i\hbar \Omega_{b} \sqrt{N_\mathrm{QW}}  
(a^\dagger -a)(\pi^\dagger_b+\pi_b)
\end{eqnarray}

Formally, the resulting Hamiltonian is equivalent to Eq.\pref{7_Hplasma}, except that the light-matter coupling constants for a single quantum well are multiplied by $\sqrt{N_\mathrm{QW}}$. This justifies our approach in part \ref{PartIII}.


\begin{thebibliography}{10}

\bibitem{Faist_QCL_1994}
J.~Faist, F.~Capasso, D.~L. ~Sivco, C. ~Sirtori, A.~L. ~Hutchinson, and A.~Y.~Cho,         
 {Sicence} {\textbf{264}}, {553-556} {(1994)}.

\bibitem{Ando_Fowler_Stern_1982}
T. Ando, A.B. Fowler, F. Stern, {Rev. of Mod. Physics}, {Vol. 52},
{No. 2}, {p. 437}, {April 1982}.

\bibitem{Dini_2003}
D. Dini, R. Kohler, A. Tredicucci, G. Biasiol, and L. Sorba,
{Phys. Rev. Lett.} {\textbf{90}}, {116401} {(2003)}.

\bibitem{Ciuti_PhysRevB_2005}
C. Ciuti, G. Bastard, and I. Carusotto, {Phys. Rev. B}
{\textbf{72}}, {115303} {(2005)}.

\bibitem{Todorov_PRL_2009}
Y.~Todorov, A.~M. Andrews, I.~Sagnes, R.~Colombelli, P.~Klang, G.~Strasser, and C.~Sirtori, {Phys. Rev. Lett.} {\textbf{102}}, {186402} {(2009)}.

\bibitem{Anappara_2009a}
A.~A.~Anappara, S.~De~Liberato, A.~Tredicucci, C.~Ciuti, G. ~Biasiol, L. ~Sorba, and F.~Beltram, {Phys. Rev. B} {\textbf{79}}, 201303(R) {(2009)}.

\bibitem{PJouy_2011}
P.~Jouy, A.~Vasanelli, Y.~Todorov, A.~Delteil, G.~Biasiol, L.~Sorba, and C.~Sirtori,
{Appl. Phys. Lett.} {\textbf{98}}, {231114} {(2011)}.

\bibitem{Geiser_2012}
M.~Geiser, F.~Castellano, G.~Scalari, M.~Beck, L.~Nevou, and J.~Faist,
{Phys. Rev. Lett.} {\textbf{108}}, {106402} {(2012)}.

\bibitem{Andreani_2003}
L.~C. Andreani in
{\textit{Proceedings of the International School of Physics Enrico
Fermi, Course CL}},  {B. Deveaud, A. Quattropani and P. Schwendimann (eds)}{(IOS Press,  Amsterdam,  2003)}.

\bibitem{Todorov_PRB_2012}
Y. Todorov and C. Sirtori, {Phys. Rev. B} {\textbf{85}}, {045304} {(2012)}.

\bibitem{Todorov_Opex_2010}
Y.~Todorov, L.~Tosetto, J.~Teissier, A.~M. Andrews, P.~Klang, R.~Colombelli,
I.~Sagnes, G.~Strasser, and C.~Sirtori, {Opt. Express} {\textbf{18}} {13886-13907} {(2010)}.

\bibitem{Book_Helm}
M. Helm in {\textit{Intersubband Transitions in Quantum Wells:
Physics and Device Applications I}} {H.C. Liu and F. Capasso eds.
} {(Academic Press, San Diego 2000)}.

\bibitem{Savona1995733}
V.~Savona, L.C.~Andreani, P.~Schwendimann, and A.~Quattropani, {Solid State Commun.} {\textbf{93}}, {733-739} {(1995)}.

\bibitem{Warburton_1998}
R.~J. Warburton, K.~Weilhammer, J.~P. Kotthaus, M.~Thomas, and H.~Kroemer,
{Phys. Rev. Lett.} {\textbf{80}}, {2185} {(1998)}.

\bibitem{Book_Sze}
S.~M. Sze.
{\textit{Semiconductor devices, physics and technology}}.
{(John Wiley \& Sons, 1985)}.

\bibitem{Todorov_PRL2010}
Y.~Todorov, A.~M. Andrews, R.~Colombelli, S.~De~Liberato, C.~Ciuti, P.~Klang,
  G.~Strasser, and C.~Sirtori,
{Phys. Rev. Lett.} {\textbf{105}}, {196402} {(2010)}

\bibitem{Todorov_JOSAA_2007}
Y.~Todorov and C.~Minot, {J. Opt. Soc. Am. A} {\textbf{24}}, {3100-3114} (2007).

\bibitem{Collin_S_2001}
S.~Collin, F.~Pardo, R.~Teissier, and J.-L.~Pelouard, {Phys. Rev. B} {\textbf{63}}, {033107} {(2001)}.

\bibitem{PJouy_APL_2011}
P.~Jouy, Y.~Todorov, A.~Vasanelli, R.~Colombelli, I.~Sagnes, and C.~Sirtori, {Appl. Phys. Lett.} {\textbf{98}}, {021105} {(2011)}.

\bibitem{Book_Haug_Koch}
H.~Haug and S.~W. Koch, {\textit{Quantum Theory of the Optical and Electronic Properties of Semiconductors}} {(World Scientific Publishing, 2004)}.

\bibitem{Wendler_Kraft_PQW_1996}
L.~Wendler and T.~Kraft, {Phys. Rev. B} {\textbf{54}}, {11436--11456} {(1996)}.

\bibitem{Zanotto_APL_2010}
S.~Zanotto, G.~Biasiol, R.~Degl’Innocenti, L.~Sorba, and A.~Tredicucci, {Appl. Phys. Lett.} {\textbf{97}}, {231123} {(2010)}.

\bibitem{note_Greg_Snider}
1D Poisson/Schr\"{o}dinger solver program developed by Dr. Gregory Snider,
  University of Notre Dame.

\bibitem{SLuin_2001}
S.~Luin, V.~Pellegrini, F.~Beltram, X.~Marcadet, and C.~Sirtori, {Phys. Rev. B} {\textbf{64}}, {041306(R)} {(2001)}.

\bibitem{Metzner_Dohler_1999}
C.~Metzner and G. H.~ Dohler, {Phys. Rev. B.} {\textbf{60}} {11005} {(1999)}.

\bibitem{Wickenbrock_2011}
A.~Wickenbrock, P.~Phoonthong, and F.~Renzoni, {J. Mod. Opt.} {\textbf{58}}, {1310–1316} {(2011)}.

\bibitem{McKeever_2003}
J.~McKeever, A.~Boca, A.~D. Boozer, J.~R. Buck, and H.~J. Kimble, {Nature}  {\textbf{425}}, {268-271} {(2003)}.

\bibitem{Ciuti_Carusotto_2006}
C.~Ciuti and I.~Carusotto, {Phys. Rev. A} {\textbf{74}}, {033811} {(2006)}.

\bibitem{Manasreh_1991}
M.O.~Manasreh, F.~Szmulowicz, T.~Vaughan, K.R.~Evans, C.E.~Stutz, and D.W.~Fischer,
{Phys. Rev. B} {\textbf{43}}, {9996} {(1991)}.

\bibitem{Dutra_Furuya_1998}
S.~M. Dutra and K.~Furuya, {Phys. Rev. A} {\textbf{57}}, {3050} {(1998)}.

\end{thebibliography}

\end{document}